\documentclass[pra,twocolumn,groupedaddress,showpacs]{revtex4}

\usepackage{epsf,latexsym}
\usepackage{times}

\newcommand\x{{\sf X}}
\newcommand\y{{\sf Y}}

\newcommand{\pr}{Phys. Rev.}

\newcommand{\ket}[1]{| #1 \rangle}

\newcommand{\ra}{{\rightarrow}}

\newcommand{\be}{\begin{equation}}
\newcommand{\ee}{\end{equation}}
\newcommand{\ignore}[1]{}

\def\CC{{\rm\kern.24em \vrule width.04em height1.46ex depth-.07ex
    \kern-.30em C}}
\def\QQ{{\rm\kern.24em \vrule width.04em height1.46ex depth-.07ex
    \kern-.30em Q}}

\def\P{{\rm I\kern-.25em P}}

\def\bbbq{{\mathchoice {\setbox0=\hbox{$\displaystyle\rm Q$}\hbox{\raise
0.15\ht0\hbox to0pt{\kern0.4\wd0\vrule height0.8\ht0\hss}\box0}}
{\setbox0=\hbox{$\textstyle\rm Q$}\hbox{\raise
0.15\ht0\hbox to0pt{\kern0.4\wd0\vrule height0.8\ht0\hss}\box0}}
{\setbox0=\hbox{$\scriptstyle\rm Q$}\hbox{\raise
0.15\ht0\hbox to0pt{\kern0.4\wd0\vrule height0.7\ht0\hss}\box0}}
{\setbox0=\hbox{$\scriptscriptstyle\rm Q$}\hbox{\raise
0.15\ht0\hbox to0pt{\kern0.4\wd0\vrule height0.7\ht0\hss}\box0}}}}

\def\bbbc{{\mathchoice {\setbox0=\hbox{$\displaystyle\rm C$}\hbox{\hbox 
to0pt{\kern0.4\wd0\vrule height0.9\ht0\hss}\box0}} 
{\setbox0=\hbox{$\textstyle\rm C$}\hbox{\hbox 
to0pt{\kern0.4\wd0\vrule height0.9\ht0\hss}\box0}} 
{\setbox0=\hbox{$\scriptstyle\rm C$}\hbox{\hbox 
to0pt{\kern0.4\wd0\vrule height0.9\ht0\hss}\box0}} 
{\setbox0=\hbox{$\scriptscriptstyle\rm C$}\hbox{\hbox 
to0pt{\kern0.4\wd0\vrule height0.9\ht0\hss}\box0}}}}

\def\bbbz{{\mathchoice {\hbox{$\sf\textstyle Z\kern-0.4em Z$}}
{\hbox{$\sf\textstyle Z\kern-0.4em Z$}}
{\hbox{$\sf\scriptstyle Z\kern-0.3em Z$}}
{\hbox{$\sf\scriptscriptstyle Z\kern-0.2em Z$}}}}

\newcommand{\putfig}[2]{$$\leavevmode\hbox{\epsfxsize=#2 cm 
\epsffile{#1.eps}}$$}

\begin{document}

\title{Quantum gates with topological phases}
\author{Radu Ionicioiu}
\affiliation{Institute for Scientific Interchange (ISI), Villa Gualino, Viale Settimio Severo 65, I-10133 Torino, Italy}
\email{radu@isiosf.isi.it}

\begin{abstract}
We investigate two models for performing topological quantum gates with the Aharonov-Bohm (AB) and Aharonov-Casher (AC) effects. Topological one- and two-qubit Abelian phases can be enacted with the AB effect using charge qubits, whereas the AC effect can be used to perform all single-qubit gates (Abelian and non-Abelian) for spin qubits. Possible experimental setups suitable for a solid state implementation are briefly discussed.
\end{abstract}
\maketitle

One of the most important problems in the field of quantum information processing (QIP) is to find a way to fight the fragility of quantum states. At present, decoherence is the main cause preventing us to implement useful (i.e., on a large number of qubits) quantum algorithms. Several methods have been proposed in order to solve this problem, including quantum error correction \cite{error_correction}, noiseless subsystems \cite{noiseless} and bang-bang control \cite{bangbang}. More recently, holonomic/geometric \cite{holonomic, geometric, solinas} and topological quantum computation (QC) \cite{kitaev, lloyd, freedman, zanardi_lloyd} have attracted an increased interest.

The main drawback of holonomic QC is the adiabaticity condition. In order to remain in the same degenerate eigenspace we need to evolve the system adiabatically. This imposes strong constraints on the time evolution of the system. There is an apparent conflict between the adiabatic/slow evolution and the requirement of performing as many gates as possible during the coherence time of the system, which led some authors to question the utility of holonomic gates for performing fault tolerant QC. In the same time there are efforts to relax the adiabaticity condition by using non-adiabatic \cite{nonadiabatic} or topological phases \cite{ericsson, pachos_vedral}.

In this article we discuss an answer to the following question: {\em Given the known topological effects, how can we use them to perform topological quantum gates?} We start with an overview of topological effects and underline their common structure. Next we construct two models which employ Aharonov-Bohm and Aharonov-Casher topological phases to perform quantum gates.
~\\
{\bf Topological phases}\\
There are four topological effects (with the corresponding phases): Aharonov-Bohm (AB) \cite{ab}, Aharonov-Casher (AC) \cite{ac} and their electromagnetic duals, the dual Aharonov-Bohm (dAB) \cite{dowling} and the He-McKellar-Wilkens (HMW) \cite{hmw}. All these four effects have a common structure: a point-like type-$\sf A$ particle moving around a linear, infinitely long distribution of type-$\sf B$ particles acquires a topological phase. In the Aharonov-Bohm effect, an electric charge $e$ moves around an impenetrable line of magnetic dipoles $\vec \mu$. In the Aharonov-Casher effect (the reciprocal of the AB effect), the roles of the two types of particles are interchanged: a magnetic moment (spin) $\vec \mu$ moves around an infinite line of charges $e$. In the dual effects, the type $\sf A$ and $\sf B$ particle are replaced by their electromagnetic duals: $e \ra g$, $\vec \mu \ra \vec d$, where $g$ is a magnetic monopole and $\vec d$ a electric dipole. It is straightforward to see that the dual effects, dAB and HMW, are also reciprocal to each other. The four effects can be summarized schematically as follows (under the electromagnetic duality mapping we also have $g \ra e$, $\vec d \ra \vec \mu$):

\[ \begin{array}{ccc}
\mbox{AB}: (e, \vec \mu) & \stackrel{\mbox{\small reciprocal}} {\longleftrightarrow} & \mbox{AC}: (\vec \mu, e) \\
\mbox{\small dual} \Big\updownarrow & & \Big\updownarrow \mbox{\small dual} \\
\mbox{dAB}: (g, \vec d) & \stackrel{\mbox{\small reciprocal}} {\longleftrightarrow} & \mbox{HMW}: (\vec d, g)
\end{array} \]
where for each effect we show the type of particles involved, e.g. $(\sf A, B)$.

All four topological effects have some important properties. Firstly, the phases are {\em non-dispersive}, i.e., they are independent of the velocity of the $\sf A$-particle moving around the line distribution of $\sf B$-particles. This relaxes the adiabaticity requirement present in holonomic implementations. Nondispersivity of both AB \cite{nondispersiveAB} and AC \cite{sangster,nondispersiveAC} effects has been verified experimentally.

Secondly, the generated phase depends only on the {\em homotopy class} of the trajectory, and not on the local details or shape of the path followed by the particle. Moreover, compared to the holonomic case, where the path should be area-preserving, here this requirement is absent. 

The main focus of this article will be on the AB and AC effects and how they can be used in quantum computing. We consider the following quantum gates: ${\sf P}(\varphi) = \mbox{diag}\,(1,\, e^{i\varphi})$ (single-qubit phase shift), ${\sf H}= 2^{-1/2}\pmatrix{1&1 \cr 1&-1}$ (Hadamard) and ${\sf C}(\varphi) = \mbox{diag}\, (1,\,1,\,1,\,e^{i\varphi})$ (controlled-phase gate). In order to perform an arbitrary unitary operation on a $n$-qubit register (i.e., achieve universality), we need to implement a universal set of quantum gates. There are several such universal sets, like $\{ \sf H, P(\varphi), C(\pi) \}$ or $\{ \sf H, C(\varphi) \}$. We will see below how we can implement some of these gates topologically. An example of a topological phase gate ${\sf P}(\varphi)$ for charge qubits (using the Aharonov-Bohm effect) has been shown in \cite{bell_test}.

The AB and AC phases have a common gauge structure. If we expand to first order Dirac equation for an electron in an external electromagnetic field, we obtain a (non-relativistic) Hamiltonian with a $U(1)_{\rm em}\times SU(2)_{\rm spin}$ gauge symmetry \cite{u1su2}. The corresponding gauge fields are exactly those giving the two topological effects: the AB and AC correspond to a $U(1)$ and $SU(2)$ phase, respectively. This suggest an obvious strategy: acting on {\em charge (spin)} degrees of freedom we obtain a $U(1)$ ($SU(2)$, respectively) topological phase.
~\\
{\bf Abelian phases: a topological lattice model}\\
We now construct a simple (quasi) topological model of a quantum register. Suppose we have two types of particles on a lattice, $\x$ and $\y$, with the following property. Whenever we move along a loop $\gamma$ an $\x$-particle around an $\y$-particle, the wavefunction picks up a phase, i.e., $\ket{\x\y} \rightarrow e^{i n\varphi}\ket{\x\y}$; $n$ is an integer defining the homotopy class of the loop, i.e., the number of times the loop $\gamma$ encircles the  $\y$-particle. Note that the phase changes the sign if we move $\y$ around $\x$. The phase is topological since it depends only on the homotopy class of the loop $\gamma$.

\begin{figure}
\putfig{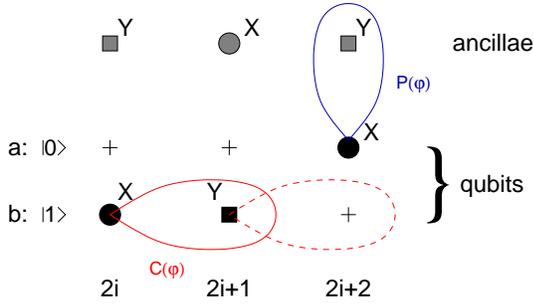}{7}
\caption{A topological lattice model; $\x(\y)$-type particles are shown as full circles (squares); qubits in black, ancillae in grey; empty spaces on the lattice are marked with '+'.}
\label{ac2}
\end{figure}

Our qubit register is an alternating array of $\x$ and $\y$ particles, such that we have an $\x\, (\y)$ particle for an even (odd) qubit index. We use a dual rail encoding: one particle in two lattice sites (or modes) labeled $a$ and $b$ (this encoding is characteristic for using ``charge'' degrees of freedom as qubit states). The qubit states are defined as follows. If the particle ($\x$ or $\y$) is in the $a$ mode, the qubit is in the logical state $\ket{0}$, and if the particle is in the $b$ mode, the qubit is in logical state $\ket{1}$ (see Fig.~\ref{ac2}). Formally, $\ket{0}\equiv a^\dagger_{\x(\y)} \ket{\rm vac}$ and $\ket{1}\equiv b^\dagger_{\x(\y)} \ket{\rm vac}$, where $a^\dagger_{\x(\y)}$ and $b^\dagger_{\x(\y)}$ are creation operators for an $\x\, (\y$) particle in $a$ and $b$ mode, respectively; $\ket{\rm vac}$ is the Fock space vacuum state. In addition, each qubit has also a fixed ancilla whose role is to enact the single qubit phase gate $\mbox{diag}\,(e^{i\varphi}, 1)= e^{i\varphi}{\sf P}(-\varphi)$. If the qubit is $\x$, its ancilla is $\y$ and vice-versa. To perform the phase gate ${\sf P}(\varphi)$, we take whatever is at the $a$-site of the qubit and move it around its ancilla (which is always in the same place). Since the phase gate is fixed (i.e., $\varphi= {\rm const}$), universality also requires $\varphi/\pi\not \in \QQ$.

To do the conditional 2-qubit phase gate ${\sf C}(\varphi)$, we take whatever is at the $b$-site of one qubit and move it around the $b$-site of the second qubit. We obtain a phase shift if and only if both particles are in the $\ket{1}$ state. In this model the only gate which cannot be done topologically is the Hadamard gate ${\sf H}$. To perform it, the particle should tunnel between the two modes (sites) $a$ and $b$ with the usual hopping Hamiltonian $H_h= \tau(t)(a^\dagger b + a b^\dagger)$, where $\tau(t)$ is the tunneling rate. The unitary evolution given by $H_h$ acting on the system during $t \in [0,\, T]$ is a single qubit rotation around the $x$-axis, i.e., ${\sf R_x}(\theta)\equiv e^{i\theta \sigma_x}$, with $\theta= -\int_0^T \tau(t) dt /\hbar$. Then the Hadamard is obtained as ${\sf H}= {\sf P}(-\pi/2)\, {\sf R_x}(\pi/4)\, {\sf P}(-\pi/2)$.

We now discuss three examples which can be cast into this framework. The first two are closely related (although not identical) to models proposed by Lloyd \cite{lloyd} and Ericsson and Sj\"oqvist \cite{ericsson}. The last one is inspired by the beautiful argument of Dirac on magnetic monopoles \cite{dirac}. Thus the topological lattice model described here can be seen as a unifying scheme for different models.~\\
{\bf (i)} $(\x, \y)= (\sf A, A)$, where $\sf A$ is an Abelian anyon. This is the Lloyd model \cite{lloyd}. An Abelian anyon moving on a loop around another anyon picks up a topological phase and all the gates follow as discussed above.\\
{\bf (ii)} $(\x, \y)= (e, \vec \mu)$; $e$ is an electric charge and $\vec \mu$ is a magnetic dipole. This is a 2D model similar to the one described in \cite{ericsson}. At first sight it seems surprising that in two (spatial) dimensions a charge moving around a spin picks up a phase. The reason is an interesting anyonization effect. As showed by Reuter \cite{reuter}, charged particles with a magnetic moment interacting via {\em standard} Maxwell electromagnetism behave like anyons in a 2+1 dimensional space-time. In 2D a point particle provides a topological obstruction (and hence a loop around it is noncontractible) in the same way as in 3D an infinite string does. Unfortunately, the scheme is strictly two-dimensional and cannot be generalized to 3D \cite{confinement}.\\
{\bf (iii)} $(\x, \y)= (e, g)$; again, $e$ is an electric charge and $g$ is a magnetic monopole. This follows from the celebrated example of Dirac, who proved that the existence of a single magnetic monopole, somewhere in the Universe, implies the quantization of all electric charges: $eg= n\hbar c/2,\ n\in \bbbz$. If the electric charge moves on a closed path around the magnetic monopole, it acquires a phase proportional to the solid angle defined by the loop (as viewed from the monopole) \cite{ryder}. For a magnetic monopole of unit magnetic charge ($eg=\hbar c/2$), the phase acquired by the electron around a loop subtending the solid angle $\Omega$ is $\varphi= \Omega/2$. Clearly, this phase is in general {\em not} topological, since is proportional to the solid angle defined by the path (from this point of view it is holonomic, with the difference that adiabaticity is not required). However, if the path is planar and the plane contains the magnetic monopole $g$, then the phase will be always $\pi$, since the solid angle is $\Omega= 2\pi$ for any planar paths with $g$ inside. Therefore, for {\em planar paths}, the phase is topological as it depends only on the homotopy class of $\gamma$. The planar geometry is well suited for the ${\sf C}(\pi)$-phase shift. The downside is that arbitrary phase gates ${\sf P}(\varphi)$ are not topological for the reason discussed above: universality requires $\varphi/\pi\not \in \QQ$ and this implies that the ancillae array should be off the qubits plane such that $\Omega/2\pi \not \in \QQ$. Thus, in this example phase gates  ${\sf P}(\varphi)$ are only holonomic. We point out that since we use charge-like degrees of freedom for the qubit definition, the influence of the magnetic field on the electron spin is not relevant in this case.

Each of these examples has some drawbacks. The first and third model work in the usual 3-dimensional space, but they require rather exotic particles, anyons and magnetic monopoles, respectively. The second model works with normal charges and spins, but only in 2+1 space-time dimensions. One may argue that model (i) is also two dimensional, since anyons exist only in two spatial dimensions \cite{wilczek}. Abelian anyons appear as quasi-particles (collective excitations) in solid-state theories of the fractional quantum Hall effect (FQHE). Abelian and non-Abelian anyons are also essential ingredients of Kitaev topological model of QC \cite{kitaev}.

Magnetic monopoles arise in spontaneously broken gauge theories in which the unbroken group $H$ is not simply connected, $\pi_1(H)\ne I$ \cite{monopole}. An example is the 't Hooft-Polyakov monopole arising in a gauge theory with a $SU(2)\ra U(1)$ spontaneous symmetry breaking (SSB). This suggests a straightforward strategy (although in practice highly nontrivial) for generating monopoles: find a SSB in a solid-state model with non-trivial $\pi_1(H)$. This will generate monopole-like configurations which can be used in the present scheme.

There is a close relationship between anyons and the AB phase. An (Abelian) anyon can be seen as a composite particle made of a point charge with a magnetic flux tube attached. Thus, when one anyon goes around another anyon, there is a nontrivial braiding of the flux tubes and the wave-function picks up a phase $e^{i\varphi}$. The origin of this phase is clearly topological: it is the AB phase. Now it is easy to see the link with statistics. Since swapping two anyons is equivalent to a half revolution of one particle around the second plus a translation, the anyon statistics is given by $\ket{2\,1}= e^{i\varphi/2} \ket{1\,2}$. From the bosons/fermions statistics ($e^{i\varphi/2}=\pm 1$), a boson or a fermion will pick up a trivial phase when one particle encircles another one. Hence, in order to have a non-trivial phase when one particle is moved around another (identical) one, we need fractional (anyonic) statistics.

It should be clear by now that in all the above examples the topological phase is equivalent to the AB phase, and hence it is Abelian. It is known that Abelian phases cannot be used alone to perform universal quantum computation and that non-Abelian phases are necessary \cite{holonomic}. We now turn to the AC phase and show how it can produce all single-qubit rotations for spin qubits.
~\\
{\bf Non-Abelian phases: spin qubits}\\
In the original AC setup \cite{ac}, a particle with magnetic moment $\vec \mu$ moving around a linear charge distribution picks up a topological phase proportional to the (linear) charge density $\lambda$ and to the homotopy class $n\in \bbbz$ of the path, $\phi_{AC}= 4\pi n\mu \lambda/\hbar c$ \cite{dowling}. This configuration (with an long line of charges) cannot be easily implemented in practice. A more appropriate setup was proposed by Casella \cite{casella} and later by Sangster {\em et al.} \cite{sangster}. The infinite line of charge can be replaced by a simple capacitor, producing a uniform electric field $\bf E$. A particle moving in the {\em static} electric field $\bf E$ will see a magnetic field ${\bf B= v \times E}/c^2$ which couples to its magnetic moment $\vec \mu$. The system is described by the spin-orbit Hamiltonian:
\be
H_{so}= \alpha\, \vec \sigma \cdot (\bf v \times E)
\label{Hso}
\ee
where $\vec \sigma$ is the vector of Pauli matrices and $\alpha= g_m e \hbar/(4m c^2)$ ($g_m$ is the gyromagnetic factor). It is straightforward to see that the phase produced by $H_{so}$ is nondispersive, $\varphi= - {\cal P}\int H_{so}\, dt/\hbar= \alpha\, {\cal P} \int_\gamma \vec \sigma \cdot ({\bf E \times dl})/\hbar$, where $\cal P$ denotes the path ordered integral.

Moreover, $\varphi$ is immune to fluctuations in velocity along $\bf E$, since ${\bf v_\parallel \times E}= 0$ (and we decomposed $\bf v= v_{\parallel}+ v_{\perp}$ in components along and perpendicular to $\bf E$, respectively). Nondispersivity also implies that $|\bf v_\perp|$ can vary, but not its {\em direction}. That is to say, once the particle has entered the gate region, it should keep the same quantization axis given by $\bf v \times E$ \cite{casella, goldhaber}. This condition is easily satisfied in a mesoscopic context if the electron moves in a 1-dimensional (1D) quantum wire perpendicular to the applied electric field $\bf E$ (e.g., using top/bottom or lateral gates). Spin rotation with {\em static} electric fields (Rashba effect) has been experimentally realized in mesoscopic heterostructures \cite{grundler, spin_orbit}.

\begin{figure}
\putfig{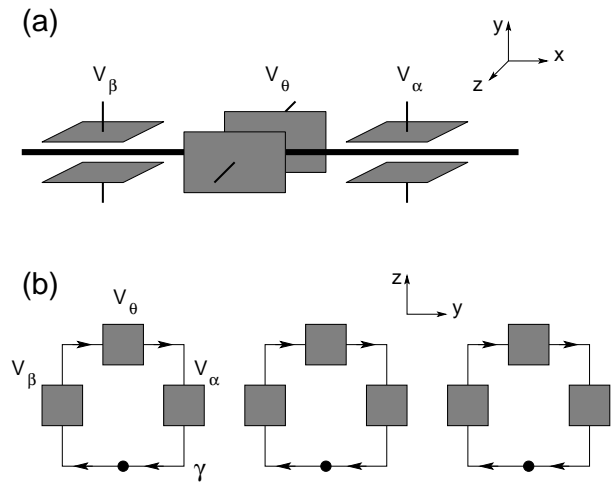}{8}
\caption{Two different architectures for a topological single qubit gate $U= e^{i\alpha\sigma_z} e^{i\theta \sigma_y} e^{i\beta\sigma_z}$.
(a) {\em Flying spin qubits}: the electron moves in a quantum wire (black line) with ${\bf v}= (v_x, 0, 0)$. Static electric fields along two directions, $E_y$ and $E_z$, produce the spin rotations ${\sf R_z}(\alpha)\equiv e^{i\alpha\sigma_z}$ and ${\sf R_y}(\theta)\equiv e^{i\theta\sigma_y}$, respectively. $V_{\alpha,\theta,\beta}$ are the gate voltages producing the phases $\alpha,\theta,\beta$; (b) {\em An array of (quasi) static spin qubits}:  $U$ is enacted by moving the spin (black dot) along the path $\gamma$. An electric field $E_x$ perpendicular to the plane of the figure is applied by top/bottom gates (grey rectangles).}
\label{su2}
\end{figure}

The Hamiltonian (\ref{Hso}) gives us the necessary ingredients to produce an arbitrary $SU(2)$ gate \cite{gates} $U(\alpha,\theta,\beta)= e^{i\alpha\sigma_z} e^{i\theta \sigma_y} e^{i\beta\sigma_z}$ by rotating the spin along two different axes \cite{aep_ri}. We discuss two possible setups for implementing spin rotations using the AC phase (see Fig.~\ref{su2}):\\
(a) the particle moves in a 1D quantum wire along the $x$-axis, ${\bf v}= (v_x, 0, 0)$. In this case we need electric fields oriented along $y$ and $z$ directions produced by top/bottom and lateral gates, Fig.~\ref{su2}a;\\
(b) if only top/bottom gates are available (e.g., the electric field is always $E_x$), $\sf R_y$ and $\sf R_z$ spin rotations can be performed by moving the particle along $z$ and $y$ directions, respectively, Fig.~\ref{su2}b.

Provided that the spin quantization axis remains the same during the gate, the phase depends only on the product between the gate length $L$ and the magnitude of the electric field, $\varphi \sim E L$. Here the product $E L$ plays the role of the linear charge density $\lambda$ in the original AC setup; the homotopy class $n$ of the path $\gamma$ is equivalent, in the Casella setup, to the number of times $\gamma$ passes through the capacitor. From an implementation point of view, this dependence is appealing, since fabrication errors in the gate length $L$ can be compensated by fine tuning the applied field $E$.

In Table \ref{tab1} we summarize the models presented above.
\begin{table}[h!]
\caption{Topological gates (and the generating effects) for the two models described.}
\begin{ruledtabular}
\begin{tabular}{l|ccr}
 & ${\sf H}$ & ${\sf P(\varphi)}$ & ${\sf C(\varphi)}$ \\
\hline
lattice model & no & yes (AB) & yes (AB) \\
spin qubits & yes (AC) & yes (AC) & no \\
\end{tabular}
\end{ruledtabular}
\label{tab1}
\end{table}

So far we investigated only how to perform the gates topologically, but we said nothing about protecting the actual qubit states. A straightforward strategy is to encode the qubits in decoherence free subspaces (DFS) and to perform topological gates on the encoded qubits. Again, in practice this could be nontrivial and implementation dependent.

In conclusion, in this article we analyzed a general framework for constructing topological quantum gates. We have shown how both Abelian and non-Abelian gates can be performed topologically using the AB and AC effects. Due to their nondispersivity, topological gates relax the (quite stringent) adiabaticity requirement present in holonomic QC. However, no universal set of topological gates has been found yet. Progress in this direction is expected by using the concept of {\em encoded universality}, i.e., a non-universal set of gates becomes universal in a suitable qubit encoding. Future work include investigating such encodings which could generate an universal set of topological gates from the (non-universal) gates presented here.

\noindent {\bf Acknowledgments.} I am grateful to Anca Popescu and Paolo Zanardi for carefully reading the manuscript and for constructive comments.

%%%%%%%%%%%%%%%%%


\begin{thebibliography}{}
\bibitem{monopole} Consider a gauge theory with a $G\ra H$ spontaneously broken symmetry ($G$ and $H$ are the symmetry groups of the Lagrangian and vacuum, respectively). The type of topological defects formed is dictated by the topology of the vacuum manifold ${\cal M}\approx G/H$. For monopoles, we need to have a nontrivial second homotopy group $\pi_2({\cal M})\ne I$. Since $\pi_2(G/H)\cong \pi_1(H)$, this implies that $H$ should be not simply-connected.
\end{thebibliography}
\end{document}